\documentclass[twocolumn,pre,a4paper,showpacs]{revtex4}
\topmargin= - 1.0cm
\usepackage{amsbsy}
\usepackage{amssymb}
\usepackage[dvips]{graphicx}

\begin{document}
\title{From simple to complex networks: inherent structures, 
barriers and valleys in the context of spin glasses}

\author{Z. Burda}
\affiliation{M. Smoluchowski Institute of Physics, Jagellonian University,
Reymonta 4, PL--30-059 Krakow, Poland}

\author{A. Krzywicki}
\affiliation{Laboratoire de Physique Th\'eorique,
b\^at. 210, Universit\'e Paris-Sud, F--91405 Orsay, France.}

\author{O. C. Martin}
\affiliation{Laboratoire de Physique Th\'eorique et Mod\`eles Statistiques,
b\^at. 100, Universit\'e Paris-Sud, F--91405 Orsay, France.}

\author{Z. Tabor}
\affiliation{Department of Biophysics, Collegium 
Medicum, Jagellonian University, PL--31-531 Krakow, Poland}

\date{\today}

\begin{abstract}
Given discrete degrees of freedom
(spins) on a graph interacting via an energy function,
what can be said about the energy
local minima and associated inherent structures? Using the lid
algorithm in the context of a spin glass energy function, we 
investigate the properties of the energy landscape 
for a variety of graph topologies. First, we find that
the multiplicity $N_s$ of the inherent structures generically has a lognormal
distribution. In addition, the large volume limit of
$\ln{\langle N_s\rangle}/\langle \ln{N_s}\rangle$ differs from unity, 
except for the Sherrington-Kirkpatrick model. 
Second, we find simple scaling laws for the growth of the
height of the energy barrier between the two 
degenerate ground states and the size of the
associated valleys. For finite connectivity models, changing
the topology of the underlying graph does not 
modify qualitatively the energy landscape, but at the quantitative level the 
models can differ substantially.
\end{abstract}

\maketitle
\vspace{1cm}
\section{Introduction}
Graphs and networks are a subject of study on their own,
and more recently the possibility of doing statistical mechanics
on these kinds of structures has been investigated. In many 
of the corresponding models,
even though the local minima of the energy function proliferate,
one knows little about their numbers or about their organisation.
Such properties are usually studied within the
``energy landscape'' paradigm which embodies
both energetic and entropic effects. In fact, a complete knowledge
of this landscape tells one
everything about low energy excitations; that kind of information
can be precious for understanding both equilibrium and
out of equilibrium low temperature properties. This is 
especially true for ``complex'' 
systems that exhibit glassy behavior; these arise in many 
subjects of research, ranging from material science to protein folding.
\par
Our purpose here is to find out how
the ``shape'' and the scaling properties
of such energy landscapes depend on the structure of the
underlying graph when the Hamiltonian function is
of the spin glass type. For this purpose, we shall consider four types
of random graphs (cf. refs. 
\cite{DominicisGoldschmidt89,VianaBray85,
KimRogers05,SherringtonKirkpatrick75}).
Previous studies have considered 
the thermodynamics of these systems: at high temperature the
system is paramagnetic, and at low temperatures there is a spin
glass phase in which the magnetizations of the spins
freeze in apparently random directions.
However relatively little attention has been
put into energy landscape questions when the 
underlying graphs are random. In particular, nearly all 
previous landscape work on our classes of graphs
has been limited to the Sherrington-Kirkpatrick 
case~\cite{SherringtonKirkpatrick75}:
the inherent structures (referred to as metastable
states in the spin-glass community) have been tackled 
analytically~\cite{BrayMoore80}, and barriers
sizes have been considered indirectly, either
via free-energy barriers near the critical 
temperature~\cite{RodgersMoore89},
or by numerical studies of 
relaxation kinetics in 
Monte Carlo~\cite{BilloireMarinari01,DallSibani03}.
Finally, there is very little work on the size of the
ground-state valleys associated with the configurations
that are reachable when staying below a given barrier.
For all of our random graph ensembles,
we shall first consider all configurations,
enumerating the inherent structures as a function of their
energy. We shall also use the 
lid algorithm~\cite{KlotzKobe94,SibaniSchriver94,SibaniPas99}
to obtain all configurations connected to the
ground state while staying below a given energy;
this gives the barrier to go from the ground state
to its inverted pair and the size of the associated basin.
In spite of the modest
graph sizes considered, a considerable
amount of information on these observables
(and in particular their scaling laws) can
be reliably extracted.

\par
The paper is organized as follows.
The models are defined in Sect.\ref{sect_MODELS}.
Then we examine all the local minima of the Hamiltonian
and study their statistics (Sect.\ref{sect_LOCALMINIMA}). 
In Sect.\ref{sect_BARRIERS} we investigate in 
detail the scaling of the energy barrier
separating the two degenerate ground states of these systems.
We also extract the size of the valley around each
ground state. We conclude in Sect.\ref{sect_CONCLUSIONS}.

\section{Models}
\label{sect_MODELS}
\subsection{Geometry: random graphs}
A first component of our models consists of a graph on whose $N$ 
vertices the spins will reside. We consider four
classes of graphs with markedly different topologies:

\par
(1) Random k-regular (KR) graphs, where the degree (connectivity) of each 
node is fixed to $k$. 
\par
(2) Erd\"os-R\'enyi (ER) graphs, where each edge is put down with probability 
$p=\alpha /N$; as a result, at large $N$, the degree of a vertex 
is a Poisson distributed variable of mean $\alpha$. 
\par
(3) Barabasi-Albert (BA) scale-free graphs generated by the usual growth 
process with the preferential attachment property \cite{BarabasiAlbert99}. 
Here, at large 
$N$, the degree distribution has a fat tail. 
In contrast to the previous graphs, the graphs in this class
are highly inhomogeneous.
\par
(4) Complete (i.e. fully connected) graphs. Here, 
the number of edges is no longer linear in $N$,
but quadratic.

\subsection{Matter: frustrated Ising spins}
To each edge $ij$ of the underlying graph, 
we independently assign a weight $J_{ij}$ 
according to a distribution symmetrized about 0, so both signs 
arise with equal probability. These elements, i.e. the random edges
and their associated weights $J_{ij}$, define the
system's ``quenched disorder''. 
The statistical mechanics problem arises when one assigns degrees 
of freedom to each site and has them interact. Here we put
an Ising spin $\sigma_{i}$ on each site $i$; the system's
Hamiltonian is taken to be
\begin{equation}
\label{H}
H(\{\sigma_{i}\}) \equiv
  -\sum_{\langle ij \rangle} J_{ij}\;\sigma_{i}\;\sigma_{j}\;,
\end{equation}
where the sum runs over all pairs of sites connected by an edge of
the graph. If not stated otherwise the weights $J_{ij}$ are generated from a
Gaussian distribution. This Hamiltonian defines a spin glass, the 
frustration coming from the fact that in general not all
terms in the energy function can be simultaneously
at their minimum. There is an obvious global $Z(2)$ symmetry
corresponding to flipping simultaneously all the spins. Finally,
because the $J_{ij}$ are i.i.d. and continuous random variables,
generically there are just two degenerate ground states (related 
by a global spin flip).

\par
We shall be interested in the large $N$ limit, in which case
it is appropriate to
keep the system's energy extensive. For instance in
the Sherrington-Kirkpatrick (SK) model~\cite{SherringtonKirkpatrick75} one sets 
$\mbox{\rm Var}[J_{ij}] = O(1/N)$. More generally, if
the mean connectivity grows, we want to keep the energy
density from diverging; in all that follows we shall thus take
\begin{equation}
\label{var}
\mbox{\rm Var}[J_{ij}] = \frac{J^2}{\langle k \rangle} ,
\end{equation}
were $\langle k \rangle$ is the average graph degree. This choice of scale
eliminates trivial differences between the values of the same observables 
in distinct models, allowing for a more direct comparison.

\begin{figure}
\includegraphics[width=5cm]{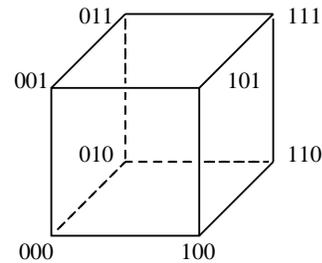}
\caption[Fig.1]{\label{fig1} Boolean (hyper) cube representing the possible
configurations of a 
three-spin system ($N=3$).}
\end{figure}

\subsection{Algorithmic procedures}
\label{subsect_algorithms}
The $2^N$ configurations of the spin glass are conveniently 
represented by vertices
of an $N$-dimensional Boolean hypercube: each vertex is identified by
a binary number $0,\dots,2^N-1$ of $N$ bits, a bit value of
1 (0) at the $i$-th 
position corresponding to the $i$-th spin being up (down). 
Neighbor vertices differ 
in exactly one bit (see fig. \ref{fig1}, where the 
case $N=3$ is illustrated). To each configuration (vertex 
of the hypercube) we associate the energy 
calculated from the Hamiltonian
(\ref{H}). The set of vertices of the 
Boolean hypercube is our configuration space; these
and their associated energies define
an energy landscape, the object 
of our study. If a vertex has an energy 
strictly lower than that of all of its
nearest neighbors, the configuration is ``one-spin-flip stable''
and hereafter will be referred to as an \emph{inherent structure}.
(As mentioned previously, such configurations are sometimes
called ``metastable states''). We use this vocabulary, 
proper to landscape studies,
in the spin-glass context. Note that no gradient algorithm (like 
those refered to when inherent structures were originally defined
on systems with continuous variables) will
be employed in this work.
\par
One instance of a problem is created by generating a random graph 
in the desired class with $N$ vertices followed by
the generation of the set of weights $J_{ij}$. The 
energy landscape of that instance
is then probed using algorithms which examine all configurations,
their energies, etc... The process is then repeated
for as many instances as possible so statistical properties
can be inferred. Finally, one studies the $N$ dependence
to extract the large $N$ scaling laws.

For $N\le 32$, a single PC machine word suffices to store the whole bit
sequence representing a configuration on the Boolean hypercube,
and furthermore many of the operations on configurations are easily 
implemented as binary operators on the corresponding machine 
words. In practice, $N \approx 30$ 
turns out to be a natural performance limit for our programs
since for larger $N$ the enumeration of all $2^N$ configurations
takes too much computation time. 

\par
The exhaustive enumeration of all inherent structures 
is in principle straightforward. 
On a 2GHz PC, it requires a few hundred seconds of CPU time for a single 
instance of $30$ spins. 
The determination of the height of the 
energy barrier separating the two degenerate ground states is more time 
consuming. We use for this purpose a variant of the 
lid algorithm\cite{SibaniSchon93}. Starting from just one of the two
degenerate ground states, one iteratively steps to 
neighboring configurations as long as their energy is below
a prescribed ``lid'' value. Using a pictorial analogy, one 
can imagine water spreading out
in a mountainous landscape: Given the source 
at the chosen ground state, the water will ``wet''
neighboring sites of the hypercube iteratively as long as their
energy is below that of the lid. Following this process,
the water front 
progresses continuously, submerging successive 
sites until a pool is formed. There exists a 
critical value of the lid beyond which the water can pour
into the basin belonging to the 
mirror ground state. This is the barrier we shall be investigating.
The computer program calculates not only the 
height of the barrier, but also the area of 
the pool (the number of wet sites) just 
before the leaking begins and the Hamming
distance from the source to the pass or
passes leading to the mirror basin.
We refer to this basin as the ground-state valley.

\par
The time needed to execute this code for one instance of 
$30$ spins 
takes from a few hundred to several thousand seconds and 
strongly depends on the geometry of the model. For example, for 
Barabasi-Albert graphs it is about 3300 seconds. This time is even 
longer for regular graphs with $k=2$: the 
ground-state valley is often very large and 
the algorithm needs much time to fill it and to reach the pass leading
to the other ground state. 

\section{Inherent structures}
\label{sect_LOCALMINIMA}
\subsection{The case of the ring geometry}
We first consider the case of the ring geometry
as it will allow us to understand the generic 
behavior of most of the other models.
We thus consider a graph that is a ring with $N$ spins
and periodic boundary conditions: this corresponds
in fact to a particular $k=2$ regular graph and it is easy
to see that the properties we shall obtain are also those
of this class of graphs (where one generally has several rings).

A configuration $\{ \sigma_i \}_{i=1,...,N}$
is an inherent structure if and only if each of
its spins is parallel to its local field. To describe
such a configuration, it is convenient to focus on
\begin{equation}
x(i,i+1) = \sigma_i J_{i,i+1} \sigma_{i+1}
\end{equation}
When $x(i,i+1)>0$, the bond $(i,i+1)$
is satisfied, otherwise it is unsatisfied. An unsatisfied bond
corresponds to having a domain wall (after resorting
to a gauge transformation). Note that because of the
periodic boundary conditions imposed, the number of
domain walls in all configurations has the same parity
as in the ground state. 

In an inherent structure, if $x(i,i+1)<0$, then necessarily
$x(i-1,i) > |x(i,i+1)| < x(i+1,i+2)$. These conditions are equivalent
to having:
(i) $|J_{i-1,i}| > |J_{i,i+1}| < |J_{i+1,i+2}|$, plus (ii)
the two neighboring bonds of the unsatisfied 
bond $(i,i+1)$ must be satisfied. Let
$M$ be the number of bonds $(i,i+1)$ for which these last two
properties hold. 
Then, there are $M$ possible binary choices (either passing 
or not passing a domain wall through each of the bonds),
leading to a number of inherent states $N_s = 2^M$,
a result derived over two decades ago~\cite{Li81,DerridaGardner86c}.
\par
For any given instance of the $J_{ij}$, one can easily
determine the set of bonds $(i,i+1)$ satisfying
$|J_{i-1,i}| > |J_{i,i+1}| < |J_{i+1,i+2}|$. Although no two
such bonds can be adjacent, the correlations are short range.
As a consequence, the number $M$ 
of these bonds is extensive as well as their variance. In
fact, $M$ is asymptotically distributed as a Gaussian
random variable according to the central limit theorem.
Furthermore, it can be shown~\cite{Li81,DerridaGardner86c} that for 
any continuous
distribution of the $J_{ij}$, one has the remarkable property
\begin{equation}
\lim_{N \to \infty} \frac{\langle M\rangle}{N} = \frac{1}{3}
\end{equation}
Since $N_s = 2^M$, we obtain
\begin{equation}
\lim_{N \to \infty} \frac{\langle \ln N_s \rangle }{N} = 
\frac{\ln 2}{3} = 0.231049...
\end{equation}
as indicated in Table~\ref{tab1}. Finally, given that $M$ is
Gaussian, $N_s$ follows a lognormal distribution at large $N$.

This simple example allows us to guess what
happens in other finite connectivity models. The role of the localized
domain wall should be replaced by a local cluster of spins
that can flip~\cite{Dasgupta79}. If there are $M$ such
clusters, the number
of inherent states will be roughly $2^M$; finally, if
as expected $M$ is a Gaussian random variable at large $N$,
then $N_s$ will be lognormal.

\subsection{The mean multiplicity}

\begin{figure}
\includegraphics[width=8cm]{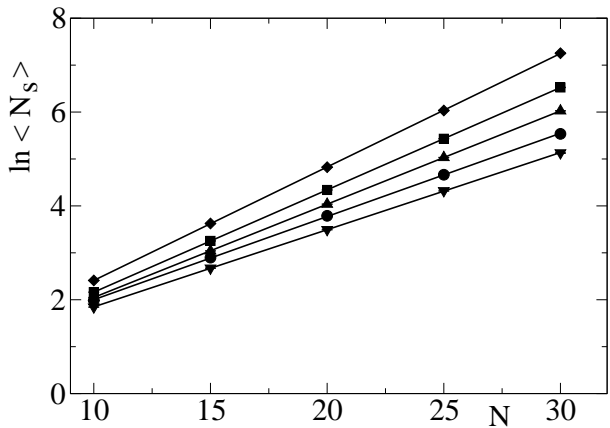}
\caption[Fig. 2]{\label{fig2} Plot of $\ln{\langle N_s\rangle}$ 
versus $N$ and associated linear fits. Diamond: KR ($k=2$), square: KR ($k=4$),
triangle up: SK,
circle: BA ($m=2$), triangle down: BA ($m=1$).}
\end{figure}

For all our models we find that the number $N_s$ of inherent 
structures grows exponentially with $N$. 
Earlier analytic 
work \cite{TanakaEdwards80,DominicisGabay80,BrayMoore80,Dean00} has 
established this for 
regular graphs (including the complete ones). We recover these 
results numerically, extending them to the 
scale-free graphs (see fig. \ref{fig2}). 
\par
The slope of $\ln{\langle N_s\rangle}$ versus $N$ is given
in Table \ref{tab1} for a sample
of models. We recall that the analytic 
result \cite{TanakaEdwards80,DominicisGabay80,BrayMoore80}
for the SK model is $0.1992\dots$, to be compared with $0.1988(2)$ read from our
table. We find this agreement
remarkable and encouraging, indicating that reliable results can be
obtained from rather small systems. We also reproduce a qualitative
result of ref.~\cite{Dean00}: for regular graphs the 
slope \emph{decreases} with increasing
connectivity. (Note that we cannot compare our 
figures quantitatively with the analytic predictions of
ref. \cite{Dean00} because the definitions 
of metastable states used are different). In contrast,
this trend with connectivity is \emph{not} found in 
inhomogeneous graphs of the Barabasi-Albert type,
where the slope is \emph{larger} for $m=2$ than for 
$m=1$ ($m$ is the number of links
attached in one step of the growth process, 
thus the average connectivity equals
$2m$ up to finite size corrections). 
Note that the case $m=1$ produces tree networks
(that is no loops are generated). 

It is also of interest to consider the average of $\ln N_s$ 
rather than the log of the average. 
The results
for these averages also show a clean linear behavior with $N$,
and the slopes are given in Table~\ref{tab1}. We see that in all
cases, the two values
are numerically close; the potential differences
will be discussed in Sect.~\ref{subsect_SK}.

\begin{center}
\begin{table}
\caption{Slope of the least-square linear fit of $\ln{\langle N_s\rangle}$
(respectively $\langle \ln{N_s}\rangle$) versus $N$. 
Only statistical errors are
estimated; the lower figure for $k=2$ is the 
exact value. The fact that for SK the lower figure is slightly larger
than the upper one is a finite-size artifact reflecting the
decrease with $N$ of the variance of $\ln{N_s}$ discussed
in the text. \label{tab1}}
\vspace{0.4cm}
\begin{center}
\begin{tabular}{|l|c|c|c|} \hline \hline
& KR $k=2$ & KR $k=4$ & ER $\langle k \rangle =4$ \\ \hline 
& & & \\
slope of $\ln{\langle N_s\rangle}$ & 0.2417(3) & 0.2179(2) & 0.2029(2)  \\ 
slope of $\langle \ln{N_s}\rangle$ & 0.2310490 & 0.2163(2) & 0.1997(2)  \\
& & & \\ \hline \hline
& BA (m=2) & BA (m=1)  & SK \\ \hline
& & & \\
slope of $\ln{\langle N_s\rangle}$ & 0.1787(3) & 0.1640(4) & 0.1988(2) \\ 
slope of $\langle \ln{N_s}\rangle$ & 0.1728(3) & 0.1510(3) & 0.2008(2) \\
& & & \\
\hline
\end{tabular}
\end{center}
\end{table}
\end{center}
\subsection{The multiplicity distribution}

The exponential growth of $N_s$ with $N$ indicates that
$\ln{N_s}$ is an extensive quantity. For finite connectivity models
it is natural to guess that this ``entropy''
arises from localized excitations~\cite{Dasgupta79} 
that are extensive in number.
Indeed, starting with a ground state, one expects 
to have an extensive number of small sized clusters (say of
just two spins as an example) that can be flipped
while keeping each spin parallel to its local field. (The
corresponding modified configuration then remains an inherent structure.)
Furthermore, this ``gas of clusters'' should
be weakly interacting.
(We thus have a generalization of what happened
in the ring geometry, although it was domain walls
that played the role of the localized objects there.)
This picture should hold for all finite
connectivity models so we focus on those first,
postponing the discussion of the SK model to later.

\begin{figure}
\includegraphics[width=8cm]{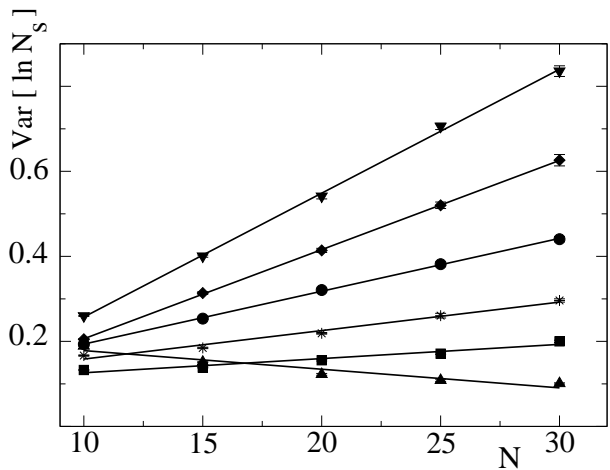}
\caption[Fig. 3]{\label{fig3} The plot of $\mbox{\rm Var}[\ln{N_s}]$ 
versus $N$ (linear regression).
Triangle down: BA ($m=1$; tree), diamond: KR $k=2$, circle: BA ($m=2$),
star: ER $\langle k\rangle =4$, square: KR $k=4$, triangle up: SK.}
\end{figure}

\begin{figure}
\vspace{1cm}
\includegraphics[width=8cm]{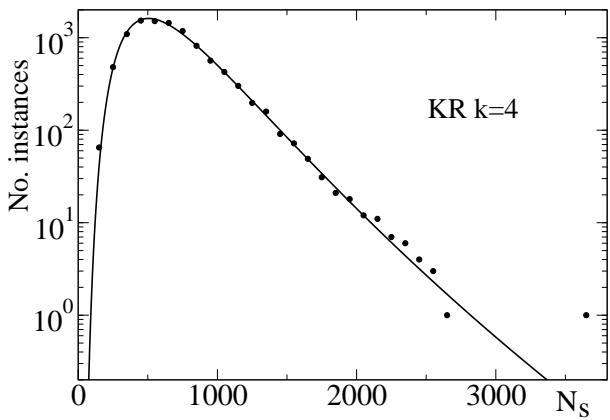}
\caption[Fig. 4]{\label{fig4} The multiplicity 
distribution for the regular $k=4$ geometry 
(histogram with bin size $=100$). The ordinate is the 
value of $N_s$ for an instance. (Summed over our instances with $N=30$,
we have a total of 10105 instances; note the
events at large $N_s$ arising from just one instance). The solid curve is 
the lognormal function with the same mean and variance as in the numerical data.}
\end{figure}

If one can excite an extensive number of
clusters and if these interact only
weakly, we expect not only the mean but also the variance
of $\ln N_s$ to be linear in $N$. Our numerical data
show that this is indeed the case as displayed in
fig. \ref{fig3}. The fits are good, especially when the slope is large.
Now pushing the weakly interacting gas picture further,
one expects a central limit
theorem behavior for $\ln N_s$. We have thus calculated its higher
order cumulants $\kappa_3$ and $\kappa_4$. Here the errors 
are rather large and so only 
the qualitative behavior with $N$ can be extracted.
The general trend is that the scaled
cumulants, i.e., $\kappa_3/\sigma^{3/2}$ and $\kappa_4/\sigma^2$, are decreasing
and probably go to zero at large $N$. (At $N=30$ they are both 
$\; {\mbox{\lower0.6ex\hbox{\vbox{\offinterlineskip
\hbox{$<$}\vskip1pt\hbox{$\sim$}}}} } \; 0.1$.)
All this indicates that the multiplicity
distribution becomes \emph{lognormal} at large $N$:
\begin{equation}
P(N_s) \simeq \frac{\exp{\bigl( -(\ln N_s - 
\langle \ln N_s\rangle)^2/2 \sigma^2\bigr)}}{N_s \sqrt{2 \pi \sigma^2}}
\end{equation}
where $\sigma^2$ is the variance of $\ln N_s$. This distribution is illustrated
for the KR $k=4$ model in fig. \ref{fig4}. Similar 
results are obtained for other models (data not shown). The $k=2$ case is 
somewhat special since the multiplicities
there are always equal to integer powers of 
$2$ as we saw in the ring geometry. 

\subsection{The special case of the SK model}
\label{subsect_SK}
Now we move on to the SK model which behaves somewhat differently from 
the other models because of its infinite connectivity.
Both $\ln \langle N_s\rangle$ and $\langle \ln N_s\rangle$
increase linearly with $N$, as already mentioned. The difference
with the other models concerns the 
variance of $\ln N_s$; as shown in fig. \ref{fig3},
this variance {\em decreases} as $N$ grows. 
(The line displayed there is to emphasize the trend,
the actual behavior is an inverse power law with $N$.) Note
that the slope of the other models is always positive
but decreases as their
connectivity increases; for instance the slope is close
to zero but definitely positive for the ER $\langle k \rangle=4$ model.

\par
Neglecting the higher order cumulants of $\ln N_s$, one has
\begin{equation}
\ln{\langle N_s \rangle} =\langle \ln{N_s} \rangle + 
\frac{1}{2} \mbox{\rm Var}(\ln{N_s})
\end{equation}
Thus, as $N \to \infty$, we have for our finite connectivity models
\begin{equation}
\frac{1}{N}[\ln{\langle N_s\rangle} - 
\langle \ln{N_s} \rangle]   \to \mbox{\rm const}
\end{equation}
This shows that in such models, the two different kinds of 
averages are distinct, though their difference is numerically
rather small.
However, for the SK model, the constant on the r.h.s. 
(right hand side) is zero because the variance is sub-extensive.
This justifies why we find 
$\langle \ln{N_s}\rangle$ to be so close to 
$\ln \langle{N_s}\rangle$, in agreement with the theoretical
result that the two averages coincide in the
thermodynamic limit. This particular property seems to be
specific to the SK model because of its infinite
connectivity.

\begin{figure}
\vspace{1cm}
\includegraphics[width=8cm]{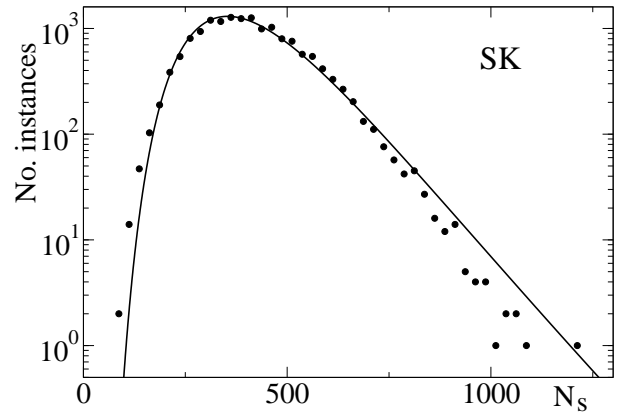}
\caption[Fig. 5]{\label{fig5} The multiplicity 
distribution for the SK model. The ordinate is an instance's
number of inherent structures (data for $N=30$ and 15607 instances). The
line is the lognormal curve with the same mean and variance 
of $\ln{N_s}$ as in the numerical data. Notice that in the tails
the data deviate somewhat from the 
curve. As stated in the text, for SK the
scaled higher order cumulants, although ``small'', are nevertheless
about twice what they are in the other models.}
\end{figure}

How can these results be understood? 
Clearly we cannot rely on the simple picture 
derived from localized clusters because in the SK model
every spin is interacting with every other one. 
Do low energy excitations nevertheless
consist of just a few spins? To find out, 
recall~\cite{Anderson78} that 
in the ground state, the field $h_i$ felt by a spin
$\sigma_i$ has a distribution 
$P(h_i) \approx | h_i |$ at small fields; thus
small fields are \emph{rare}. Because of this, the
number of two spin excitations that will
produce an inherent structure is only
$O(1)$! Going from 2 to $k$ spins increases
this number, in fact it becomes exponentially large in $k$.
Because of this, one is driven to $k=O(N)$ so the
vast majority of inherent structures correspond to 
excitations with $O(N)$ flipped spins, explaining why
$\ln N_s$ is extensive in the SK model. Furthermore, for 
such large number of spin flips, the detail of the 
$J_{ij}$ gets washed out. Thus one expects very little
instance to instance fluctuations of $\ln N_s$, suggesting
correctly that $\mbox{\rm Var}(\ln{N_s})$ is small.

Finally, we have investigated the distribution 
of $\ln N_s$ (see fig. \ref{fig5}).
Surprisingly, just as in the finite connectivity models,
it seems to be lognormal, as indicated
by the smallness of cumulants of order 3 and 4; these
cumulants are rapidly decreasing
with $N$ and are around $0.2$ at $N=30$.
We have no qualitative justification for this 
simple result.

\subsection{The multiplicity as a function of energy}
To get further insight, it is instructive to consider
the dependence of the number of inherent structures
as a function of their energy. We first define 
the scaled excitation energy per spin
$\varepsilon = (E - E_0)/JN$ where $E_0$ denotes 
the ground state energy. Second, we introduce a binning for
$\varepsilon$ and define $D(\varepsilon)$ as the number of
inherent structures whose (excitation) energy density is in 
$\left[\varepsilon, \varepsilon+ \delta \varepsilon \right]$,
divided by $\delta \varepsilon$. 

Actually, it is convenient to renormalize $D(\varepsilon)$ by dividing
it further by $\sqrt{N}$, as we now explain.
Let $s(\varepsilon)$ denote the density of 
the ``configurational entropy'' so that
\begin{equation}
\langle N_s \rangle = \int d\varepsilon \langle D(\varepsilon)\rangle
=  \int d\varepsilon \exp{\bigl( Ns(\varepsilon) \bigr)}
\label{Dvss}
\end{equation}
Assume that $s(\varepsilon)$ takes its maximum value at 
$\varepsilon=\varepsilon_m$. Hence at large $N$
\begin{eqnarray}
\langle N_s \rangle \propto \int d\varepsilon \exp{\bigl( N[s_0 + 
\frac{1}{2}s_2(\varepsilon-\varepsilon_m)^2]\bigr)} \nonumber \\
\propto \exp{(Ns_0)}/\sqrt{N}
\end{eqnarray}
However, our data, especially for the SK model, indicate that $N_s$
increases exponentially with $N$, without any power prefactor. If so,
there must be 
a factor $\sqrt{N}$ missing in (\ref{Dvss}), in the relation 
between $D(\varepsilon)$ and $\exp{\bigl(Ns(\varepsilon)\bigr)}$. Hence, 
we {\em redefine}
$s(\varepsilon)$ by 
$s(\varepsilon) = N^{-1}\ln{[\langle D(\varepsilon)\rangle/\sqrt{N}]}$, 
which amounts to adding a finite-size correction. To simplify
the writing we hereafter absorb the factor 
$1/\sqrt{N}$ in the definition of $D(\varepsilon)$. 
\par
\begin{figure}
\includegraphics[width=8cm]{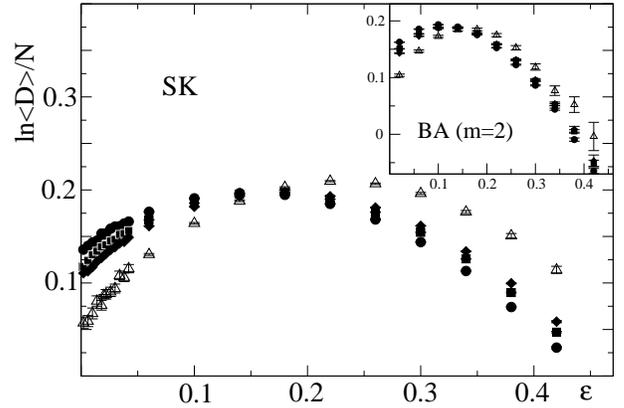}
\caption[Fig.6]{\label{fig6} The histogram 
of $N^{-1}\ln{\langle D(\varepsilon)\rangle}$ versus 
the excitation energy per spin $\varepsilon $ for 
the SK model at $N=20$ (circles), $25$ (squares) and $30$ (diamonds). The
triangles are the result of an $N \to \infty$ extrapolation. The
bin size is $0.004$ for $\varepsilon < 0.04$ and $0.04$ otherwise.
Inset: the same for the Barabasi-Albert geometry, $m=2$.
}
\end{figure}

In fig. \ref{fig6} we show 
$s(\varepsilon) = N^{-1}\ln{\langle D(\varepsilon)\rangle}$ 
versus $\varepsilon$ for 
the SK model (and in the inset for the BA (m=2) model) for
$N= 20, 25, 30$ as well as our $N=\infty$ extrapolation. The result of
this extrapolation is close to that shown in fig. 2 of 
ref. \cite{BrayMoore80}, where it has been 
calculated in the mean field approximation, 
with two differences: it does 
not fall to zero when $\varepsilon \to 0$, and at the maximum it
overshoots by $4.5\%$ the exact value $0.1992\dots$. 
However, our data have been collected at small $N$, 
and there are uncertainties with our extrapolation, so
the agreement is actually pretty good. For 
the SK model, the extrapolated data are very well fitted by the parabola
\begin{equation}
s(\varepsilon) = 0.064167 + 1.2582 \varepsilon - 2.7173 \varepsilon^2
\end{equation}
This has a maximum at $\varepsilon \approx 0.23$ and vanishes at
$\varepsilon \approx 0.51$. 

\begin{figure}
\includegraphics[width=8cm]{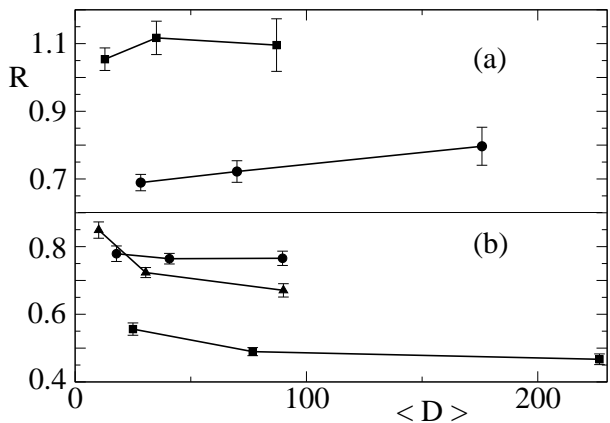}
\caption[Fig.7]{\label{fig7} The ratio 
$R=\sqrt{\langle D^2\rangle - \langle D\rangle^2}/\langle D\rangle$ 
versus $\langle D\rangle$. (a) Barabasi-Albert geometry,
circles : $0 < \varepsilon < 0.16$, squares: 
$0.16 < \varepsilon < 0.32$ (b) SK geometry, 
circles: $0 < \varepsilon < 0.12$, squares:  
$0.12 < \varepsilon < 0.24$, triangles: 
$0.24 < \varepsilon < 0.36$.}
\end{figure}

An interesting quantity (independent of the normalization of
$D(\varepsilon)$) is the ratio
\begin{equation}
R(\varepsilon) = \frac{\sqrt{\mbox{\rm Var}[D(\varepsilon)]}} 
{\langle D(\varepsilon)\rangle}
\label{ratio}
\end{equation}
It is plotted in fig. \ref{fig7}. In part (a) of that figure, 
we illustrate the behavior of $R(\varepsilon)$
for the Barabasi-Albert geometry; the ratio 
is either constant or increases slowly
with $\langle D(\varepsilon)\rangle$ (and thus with $N$). A 
similar behavior is observed for other finite
connectivity models. In part (b), we show 
$R(\varepsilon)$ for the SK model; the ratio
is constant for the energy bin 
$[0,0.12]$ and {\em decreasing} for larger
excitation energies.
\par
We also find that for all $\varepsilon$ the shape of the distribution of 
$D(\varepsilon)$ is consistent with a lognormal law.
If it were exactly lognormal, one would have
\begin{equation}
R(\varepsilon) = \sqrt{\exp{\bigl( \sigma^2(\varepsilon)\bigr)} - 1} 
\end{equation}
where $\sigma^2(\varepsilon)=\mbox{\rm Var}[\ln{D(\varepsilon)}]$. 
\par
If $R$ keeps decreasing as 
$\langle D\rangle \to \infty$, the distribution 
of the density of inherent structures 
becomes more and more peaked, leading to
\begin{equation}
\lim_{N \to \infty} 
\ln{\langle D(\varepsilon)\rangle} / \langle \ln{D(\varepsilon)}\rangle
= 1
\label{asymp}
\end{equation}
This is what Bray and Moore find in the SK model for scaled 
excitation energies $> 0.12$, where as they 
claim the ``metastable states are uncorrelated''
\cite{BrayMoore80}. In our data the fall of $R$ is not 
as rapid as expected for a Poisson distribution, 
but the qualitative trend is similar. Unfortunately,
we are unable to determine the critical energy very
precisely, we can only state that our SK data  
are compatible with the results of ref. \cite{BrayMoore80}. 

\section{Barriers and valleys}
\label{sect_BARRIERS}
\subsection{Framework}
The motivation for studying the scaling behavior of energy barriers separating
inherent structures is well known: barriers that grow with $N$ suggest a frozen 
(spin glass phase) at $T=0$ while finite barriers suggest 
a paramagnetic system.
Here we focus on the energy barrier between the two ground states related 
by spin flip symmetry, which is expected to be the system's largest barrier. 
To go from one ground state to the other, all spins must be flipped and so 
we are interested in finding:
\par
(a) the height of this barrier; 
\par
(b) the appropriately defined distance between the ground state 
and this barrier state; 
\par
(c) the number of configurations that can be reached starting from one
ground state while staying below this barrier height.
\par

\subsection{The barrier exponent}

\begin{center}
\begin{table}
\caption{The barrier and the valley exponents. Top numbers:
slope of the least-square linear fit of $\ln B$
versus $\ln{N}$. Bottom numbers: slope of
$\ln{\ln{N_{\mbox{\rm wet}}}}$ versus $\ln{N}$. 
Only statistical errors are estimated. \label{tab2}}
\vspace{0.4cm}
\begin{center}
\begin{tabular}{|l|c|c|c|c|} \hline \hline
& ER $\langle k \rangle =4$ & KR $k=4$ & BA (m=2)  & SK \\ \hline
& & & & \\
barrier exponent & 0.285(4) & 0.244(6) & 0.363(4) & 0.335(3) \\ 
valley exponent & 0.781(5) & 0.708(5) & 0.811(4) & 0.644(6)\\
& & & & \\
\hline
\end{tabular}
\end{center}
\end{table}
\end{center}
\par
The energy landscape of
a disordered and frustrated system generically has many
valleys and energy barriers. Studies of barriers in spin glasses
have focused almost exclusively on the 
Sherrington-Kirkpatrick case: (1) On the analytical side,
Rodgers and Moore~\cite{RodgersMoore89} performed
an analysis of free-energy barriers near $T_c$
and found an $N^{1/3}$ scaling. (2) Numerical investigations
have estimated barriers \emph{indirectly} via the relaxation
times of Monte Carlo dynamics; the most recent 
simulations~\cite{BilloireMarinari01} give further support to an
$N^{1/3}$ scaling. Very recently~\cite{MontanariSemerjian05}, 
a detailed study of barriers
has been performed analytically for spin glasses on
random graphs, but for a Hamiltonian involving
three-spin interactions rather than the 
two-spin interactions of Eq.~\ref{H}.
In this work we shall study directly the energy barrier scalings
for our four kinds of models, finding that the exponent depends
on the nature of the underlying graph.

The energy barrier of interest is the one encountered
when going from one ground state to its flipped counter part.
(Recall that our system has a global spin flip symmetry.)
This energy barrier is expected to be the largest of all barriers. We
determine it with our lid algorithm by letting the ``water'' proceed
from the source (one of the ground states) up to the
level given by the lid. When this level crosses the value
$B_J$ (the barrier for the instance under consideration),
the water will flow all the way to the other ground state.
$B_J$ is a random variable (depends on the instance), so we have
determined its moments and distribution as a function of $N$.

\par
\begin{figure}
\vspace{1.5cm}
\includegraphics[width=8cm]{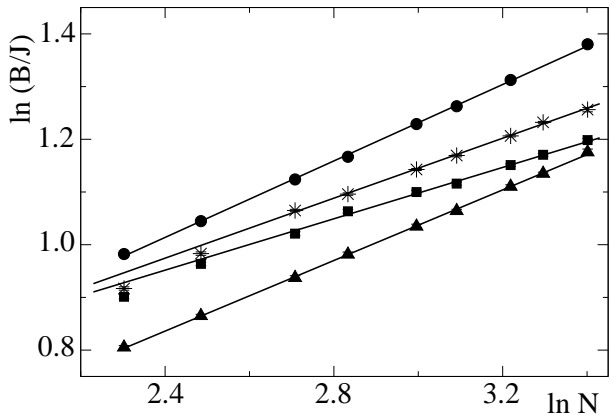}
\caption[Fig. 8]{\label{fig8} The barrier height $B$ as a function of $N$:
circle: BA ($m=2$), star: ER ($\langle k\rangle=4$), 
square: KR ($k=4$), triangle up: SK.
The ordinate for the SK points has been shifted by $-0.05$ to make the figure
less cluttered. }
\end{figure}

For each $N$ and kind of model (KR, ER, ...), we have a numerical
estimate of the distribution of $B_J$; define $B$ as the energy 
for which this distribution is maximum.
We show in fig. \ref{fig8} the log-log plots of $B$ versus $N$ for 
several of our models.
(A similar plot can be obtained from the mean of $B_J$, but our
definition of $B$ leads to more
stable results). The numerical values of the slopes 
are collected in Table \ref{tab2}. 
\par
The errors given in Table~\ref{tab2} are statistical only.
The systematic errors are likely to be more important. We do not
have them fully under control, but they can be roughly estimated
as follows: the value of the exponent depends on the observable
used to extract it, namely the average of the barrier height
or its modal value. Comparing these different observables, we
expect the systematic error on the barrier exponents
to be around $\pm 0.02$. Similarly, the systematic error
attached to the valley exponents discussed in 
the next section is tentatively estimated to
be around $\pm 0.05$. 
\par
The result for the 
SK model gives strong support for the conjecture
of a barrier scaling as $N^{1/3}$; note how well
the data follow this scaling, starting even at such
low values as $N=10$. Power scaling is also
very clear in the other models, though the scalings
seem to set in a bit less early. For instance, our fits
for KR graphs with $k=4$ lead to 
effective slopes that decrease as the lower range in $N$ used
for the fitting is increased; our
best estimate for the exponent is then $0.244(6)$, 
a result completely incompatible with $1/3$. We thus conclude
that barrier exponents depend on the nature of the underlying graphs.
It is manifest that for 
the homogeneous finite connectivity models studied here the 
barrier exponent is significantly lower than for SK (a 
similar result is found e.g. for Erd\"os-R\'enyi graphs 
with average degree equal to $3$). The inhomogeneous 
Barabasi-Albert models behave differently, more like SK, 
and there is no indication in our data that such a trend
is due to finite size effects, although this possibility 
cannot, of course, be totally excluded.
\par
In contrast to the quantity $N_s$, we find that $B_J$ is self-averaging,
i.e. its relative fluctuations \emph{decrease} and
go to zero as $N$ grows: 
\begin{equation}
\frac{\sqrt{\mbox{\rm Var}[B_J]}}{\langle B_J\rangle} \propto N^{-\gamma}
\end{equation}
with the exponent $\gamma$ ranging from about $0.14$ (SK model) to
$0.30$ (KR $k=4$ model). 
From such a self-averaging behavior,
one may guess that $B_J$ is to some extent a sum of many
small barrier heights.

\vspace{0.5cm}

\subsection{The valley exponent}
Let us consider now the number of configurations
that are wet just before the
barrier is reached. Since the procedure wets all reachable 
configurations in its search of
the pass, this number measures the size of the valley 
or, stated differently, the
size of the basin of attraction of the ground state. It 
turns out to depend strongly on the geometry 
of the model. Thus, for example, at $N=30$, 
the average number of wet sites
is 2940 in the SK and 40000 in 
the BA ($m=2$) model, respectively!
\par
Each wet site represents a configuration of the system. We observe that 
\emph{nearly all} configurations or their associated
flipped partners whose excitation energies are lower than $B_J$, the height 
of the barrier, get wet. This is true for all models 
and says that $B_J$ is probably as expected the largest
energy barrier in the system.
\par
We define $N_{\mbox{\rm wet}}$ as the most probable number of wet 
configurations when the lid finally reaches the barrier.
(It is thus extracted from all of our instances at a given $N$,
just as $B$ was.) We find the simple scaling property:
\begin{equation}
N_{\mbox{\rm wet}} \propto \exp{(\mbox{\small \rm c} N^\nu)}
\label{wet}
\end{equation}
where $c$ is a numerical constant.
The value of the ``valley exponent'' $\nu$ is always below 1,
see Table~\ref{tab2}. (Note that we have also
performed the analysis using the mean number of wet configurations
rather than the modal value and the results are very similar.)

\par
Let us present a heuristic argument for why a stretched
exponential as in Eq.(\ref{wet}) is natural.
The wet sites represent 
configurations with energies differing relatively little
from the ground state energy $E_0$, the extreme energy of the system, 
proportional to $N$. A simple guess is that the (microcanonical)
entropy density $s(e)$ vanishes as 
\begin{equation}
s(e) \sim A (e - e_0)^{\alpha} 
\end{equation}
as $e=E/N$ tends towards the ground
state energy density $e_0=E_0/N$. (Here $A$ is a constant
and $\alpha$ is a positive exponent.) 
Then the probability that a configuration has energy density $e$ is
\begin{equation}
P(e) \propto \exp{\bigl( N A (e - e_0)^{\alpha}\bigr)}
\label{tail}
\end{equation}
for $e$ tending towards $e_0$. Using the fact that for configurations
below the energy barrier $B_J$, $e-e_0$ becomes infinitesimal
as $N\to\infty$, the number of wet configurations is
\begin{equation}
N_{\mbox{\rm wet}} \propto \int_{e_0}^{e_0+B/N} P(e) de  
\propto \int_0^{N^\beta} \exp{\bigl( \mbox{\small
\rm A} N (z/N)^\alpha\bigr)} dz
\end{equation}
where $\beta$ is the barrier exponent. For large $N$ the integral is dominated
by the upper limit and one gets (\ref{wet}) with $\nu=\alpha \beta -\alpha+1$.
Using our measurements of $\beta$ and $\nu$, this gives 
values of $\alpha$ ranging from about $0.3$ to approximately $0.5$.
Related considerations can be found in ref.~\cite{KlotzSchubertHoffmann98}.
\begin{figure}
\vspace{1cm}
\includegraphics[width=8cm]{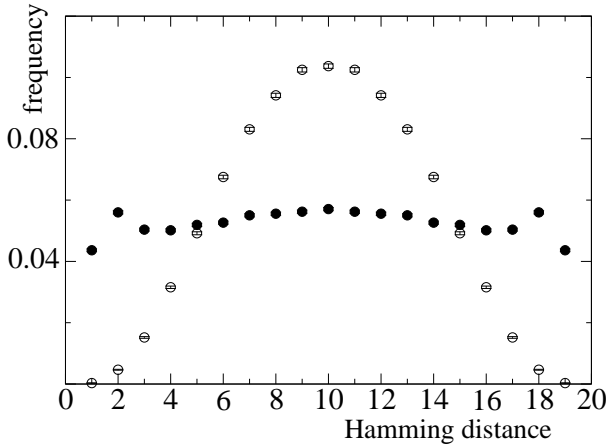}
\caption[Fig. 9]{\label{fig9} The normalized Hamming 
distance distribution separating the Boolean hypercube sites 
representing the ground
state and one or the other saddle-point (for $N=20$); 
empty circles: SK, full circles: KR $k =3$. }
\end{figure}

\subsection{Ground state to barrier state distances}
\label{subsect:Distances}
The idea of our algorithm was briefly explained in 
Sect.~\ref{subsect_algorithms}. 
Along the shortest path from the ground state
to the saddle point (pass), a given spin can get flipped more than 
once. This path has to avoid multiple obstacles and can be 
quite tortuous. It is not explicitly constructed by our simple 
``wetting'' algorithm which merely identifies
the location of the saddle point in the configuration 
space. Once this is done, one easily finds the
Hamming distance separating the ground state and the 
saddle point, i.e. the barrier state. The
mirror state is then also easily identified. 
\par
The distribution of the Hamming distances separating the ground state from
the pass is shown in fig.~\ref{fig9}: the distribution is the least broad
for the SK model and the broadest for regular graphs with $k=3$. Other
models, including the Barabasi-Albert one, exhibit intermediate behaviors.
\par
Detailed investigations shows that 
just before the lid energy is reached, in almost all cases
the two passes (related by a global spin flip) are \emph{both}
on the boundary of the wet sites. From this fact,
we can reach a useful qualitative picture of
the energy landscape. Recall that each ground
state valley corresponds to a connected cluster of 
vertices of the $N$ dimensional Boolean hypercube. Generically, two 
barrier configurations, related by a global spin flip
symmetry, form the ``passes'' where the two valleys ``touch''.
This is very different from what happens in a one dimensional
energy landscape. There, if one has two passes, then
the two valleys do not touch, instead
an intermediate zone separates them.
In our systems, only on very rare occasions (at the level of about 1 in 
1000) have we found that a valley is in contact with just one pass.
Thus in the generic case, the valley to valley barrier
is associated with two passes, both in contact 
with each ground state valley.

\section{Summary and conclusion}
\label{sect_CONCLUSIONS}

Many complex systems can be represented by a complicated network
connecting a large number of simpler subunits. The knowledge of the 
network architecture is an important piece of information, but it 
does not tell much about the system's global behavior. Instead,
it is frequently the 
energy landscape that is relevant for the cooperative behavior
of systems consisting of many interacting units (see refs.
\cite{StillingerWeber84,Fraunfelder97}). An interesting question
is: what is the relation 
between the system's behavior and its design? Between the shape of its 
energy landscape and the topology of its network? This question
motivated our investigation.
\par
Our study focuses on spin glasses that are archetypes 
of complex systems. A major conclusion of our work
is that reliable results can be obtained from quite
small systems, provided the enumeration of configurations is 
exhaustive: for example,
in the SK model we reproduce correctly the first three significant digits  of
the quantity $N^{-1} \ln{ \langle N_s \rangle}$ (which has 
been calculated analytically long
ago). The techniques we use are relatively straightforward,
but allow us to unveil essential aspects
of the energy landscape in these systems as we now summarize.
\par
We find that the distribution of the number of inherent structures 
is lognormal in all examined models. This usually reflects the extensivity
of local excitations and their low level of
correlations, a property that should hold in a
variety of other models with finite connectivity.
\par
We further observe a simple scaling behavior 
with the number of spins of the
size of the ground state basin of attraction 
and of the height of the energy barrier between the two 
degenerate ground states, respectively. However, the corresponding
exponents are not universal. For instance, the barrier
exponent in the SK model (equal to $1/3$) is definitely larger than
the exponents found in all models where underlying graphs are
homogeneous. Similar results hold for the valley exponent.
Nevertheless there is some level of universality:
the exponents seem to be independent
of the underlying $J_{ij}$ distribution. For 
instance in the SK model, the barrier exponent is,
within the limits of our systematic error, the same when the $J_{ij}$ 
are generated from an exponential distribution and
from the Gaussian distribution; we have also checked this
for the KR $k=4$ model, though finite size effects there were
stronger. We find this level of universality to arise
also for the valley exponent.
\par
On the whole, the qualitative behavior of the models we
considered is very similar,
with one notable exception: the variance of $\ln{N_s}$ \emph{decreases}
with $N$ in the SK model, while it \emph{increases}
in all models with finite connectivity graphs. This difference
is quite essential and signals a qualitatively different
nature of this complex system, arguably due to the
absence of small scale excitations.
\par
In spite of their qualitatively similar behavior, the 
models are very different
quantitatively. The seemingly small differences between parameters plotted
in figs. \ref{fig2}-\ref{fig3} translate into large differences of multiplicity
distributions. For instance, the 
size of the basin of attraction in the BA ($m=2$) model is
one order of magnitude larger than in the model
where graphs are regular with degree $4$, etc.
\par
This paper is an exploratory one and many
questions remain open. We showed in Sect.~\ref{subsect:Distances}
that the two ground state valleys had a connectivity property
that was inexistent in one dimensional energy landscapes;
it would be of interest to understand the nature of passes
between more general valleys in these systems. It would also
be worthwhile to explore
the topology of the graph associated with
the inherent structures. Such {\em inherent networks} have
been constructed, for
example, for certain atomic clusters \cite{Doye02,DoyeMassen05}. 
We would like to
see whether there is some relation between the underlying
topology of the model and the topology of its 
inherent network. However, the nontrivial definition of the latter
for a discrete system like a spin glass requires a 
separate discussion; this issue is
beyond the scope of the present paper.

\par\begin{center}
{\bf  \small ACKNOWLEDGMENTS}
\end{center}
We thank S. Majumdar for stimulating discussions and P.L. Krapivsky
and A. Montanari for comments.
We are also indebted to B. Waclaw for his 
participation in the early stage of this project.
This work was supported by the EEC's FP6 Information
Society Technologies Programme
under contract IST-001935, EVERGROW (www.evergrow.org), by the
EEC's HPP under contract HPRN-CT-2002-00307 (DYGLAGEMEM),
by the EEC's FP6 Marie Curie RTN under
contract MRTN-CT-2004-005616 (ENRAGE: European
Network on Random Geometry) and by the Polish
MSIST grants 2P03B-08225 (2003-2006) and
1P03B-04029 (2005-2008).
The LPT and LPTMS are Unit\'e de Recherche de
l'Universit\'e Paris~XI associ\'ee au CNRS.


\bibliographystyle{prsty}

\bibliography{references}

\addcontentsline{toc}{chapter}{\protect\bibname}
\begin{thebibliography}{10}

\bibitem{DominicisGoldschmidt89}
C. {De Dominicis} and Y. Goldschmidt, J. Phys. A {\bf 22},  L775  (1989).

\bibitem{VianaBray85}
L. Viana and A.~J. Bray, J. Phys. C {\bf 18},  3037  (1985).

\bibitem{KimRogers05}
D.H. Kim, G.J. Rodgers, B. Kahng, and D. Kim, Phys. Rev. E {\bf 71},  056115
  (2005).

\bibitem{SherringtonKirkpatrick75}
D. Sherrington and S. Kirkpatrick, Phys. Rev. Lett. {\bf 35},  1792  (1975).

\bibitem{BrayMoore80}
A.~J. Bray and M.~A. Moore, J. Phys. C {\bf 13},  L469  (1980).

\bibitem{RodgersMoore89}
G. Rodgers and M. Moore, J. Phys. A {\bf 22},  1085  (1989).

\bibitem{BilloireMarinari01}
A. Billoire and E. Marinari, J. Phys. A {\bf 34},  L727  (2001).

\bibitem{DallSibani03}
J. Dall and P. Sibani, Eur. Phys. J. B {\bf 36}, 233 (2003).

\bibitem{KlotzKobe94}
T. Klotz and S. Kobe, J. Phys. A {\bf 27},  L95  (1994).

\bibitem{SibaniSchriver94}
P. Sibani and P. Schriver, Phys. Rev. B {\bf 49},  6667  (1994).

\bibitem{SibaniPas99}
P. Sibani, R. van~der Pas, and J. Schoen, Comp. Phys. Comm. {\bf 116},  17
  (1999).

\bibitem{BarabasiAlbert99}
L. Barabasi and R. Albert, Science {\bf 286},  509  (1999).

\bibitem{SibaniSchon93}
P. Sibani, J.~C. Sch{\"o}n, P. Salamon, and J.~O. Andersson, Europhysics
  Letters {\bf 22},  479  (1993).

\bibitem{Li81}
T. Li, Phys. Rev. B {\bf 24}, 6579 (1981).

\bibitem{DerridaGardner86c}
B. Derrida and E. Gardner, J. Physique {\bf 47}, 959 (1986).

\bibitem{Dasgupta79}
C. Dasgupta, S.K. Ma, and C.K. Hu, Phys. Rev. B {\bf 20}, 3837 (1979).

\bibitem{TanakaEdwards80}
F. Tanaka and S. Edwards, J. Phys. F {\bf 10},  2769  (1980).

\bibitem{DominicisGabay80}
C. {De Dominicis}, M. Gabay, T. Garel, and H. Orland, J. Physique {\bf 41},
  923  (1980).

\bibitem{Dean00}
D.~S. Dean, Eur. Phys. J. B {\bf 15},  493  (2000).

\bibitem{Anderson78}
P.~W. Anderson,  in {\em Ill-condensed matter, Proceedings of the Les 
Houches Summer School of Theoretical Physics, session XXXI (1978)}, 
ed. R. Balian et~al (North Holland, Amsterdam, 1979).

\bibitem{MontanariSemerjian05}
A. Montanari and G. Semerjian, Phys. Rev. Lett. {\bf 94},  247201  (2005).

\bibitem{KlotzSchubertHoffmann98}
T. Klotz, S. Schubert and K.H. Hoffmann, Eur. Phys. J. B {\bf 2}, 313 (1998). 

\bibitem{StillingerWeber84}
F.~H. {Stillinger} and T.~A. {Weber}, Science {\bf 225},  983  (1984).

\bibitem{Fraunfelder97}
 {\em {Procedings of 16th Annual Int. Conf. of the Center for Nonlinear Studies
  (Los Alamos)}}, ed. H. Fraunfelder et~al., 
  Physica D {\bf 107}, 117-439 (1997).

\bibitem{Doye02}
J.P.K. Doye, Phys. Rev. Lett. {\bf 88},  238701  (2001).

\bibitem{DoyeMassen05}
J.P.K. Doye and C.P. Massen, Phys. Rev. E {\bf 71},  016128  (2005).

\end{thebibliography}

\end{document}